\documentclass[twocolumn,aps,showpacs]{revtex4}
\usepackage{amsfonts}
\usepackage{amsmath}
\usepackage{graphicx}
\usepackage{bm}
\usepackage{amssymb}
\usepackage{dcolumn}

\begin{document}

\title{Effects of the crystal structure on the ferromagnetic correlations \\
in ZnO with magnetic impurities}

\author{Bo Gu$^{1}$, Nejat Bulut$^{1,2}$, and Sadamichi Maekawa$^{1,2}$}

\affiliation{$^1$ Institute for Materials Research, Tohoku
University, Sendai 980-8577, Japan\\ $^{2}$CREST, Japan Science and
Technology Agency (JST), Kawaguchi, Saitama 332-0012, Japan}

\begin{abstract}
We study the ferromagnetism in the compound (Zn,Mn)O within the
Haldane-Anderson impurity model by using the quantum Monte Carlo
technique and the tight-binding approximation for determining the
host band-structure and the impurity-host hybridization. This
computational approach allows us to determine how the host crystal
structure influences the impurity bound state, which plays an
important role in the development of the ferromagnetic (FM)
correlations between the impurities. We find that the FM
correlations are strongly influenced by the crystal structure. In
particular, in p-type (Zn,Mn)O, we observe the development of FM
correlations with an extended range at low temperatures for
wurtzite and zincblende crystal structures. However, for the
rocksalt structure no FM correlations are observed between the
impurities. In addition, in n-type ZnO with magnetic impurities,
the impurity bound state and FM correlations are not found.
\end{abstract}

 \maketitle

\section{Introduction}

Dilute magnetic semiconductors (DMS) could lead to new spintronic
devices where both the electronic charge and spin can be
controlled.  For practical applications, DMS with Curie
temperature, $T_c$, above room temperature are required. The
(Ga,Mn)As is regarded as a classic example with robust
ferromagnetism \cite{GaMnAs1}, but the highest reported $T_c$'s
are still well below room temperature \cite{GaMnAs2, GaMnAs3}.
Alternatively, p-type (Zn,Mn)O has been predicted to be a room
temperature ferromagnet \cite{ZnMnO1, ZnMnO2}, where
ferromagnetism above room temperature has been observed for Mn
doped pure ZnO \cite{InRoomT1,InRoomT2,InRoomT3}, or Mn doped
p-type ZnO \cite{PRoomT1,PRoomT2,PRoomT3,PRoomT4,PRoomT5}.
However, contradictory experimental results have also been
reported for (Zn,Mn)O such as ferromagnetism below room
temperature \cite{LowT}, absence of ferromagnetism \cite{NoFM},
spin glass behavior \cite{Glass}, or paramagnetism \cite{Paramag}.

The possibility of room-temperature ferromagnetism motivates the
theoretical research on the origin and control of high-temperature
ferromagnetism in semiconductors, especially ZnO-based DMS, which
is the subject of this paper. Magnetic impurities in ordinary
metals exhibit the well-known Ruderman-Kittel-Kasuya-Yosida (RKKY)
oscillations, which is a carrier mediated indirect coupling due to
the Friedel oscillations of the polarized carriers around the
impurities. When magnetic impurities are doped into a
semiconductor host, quite different behaviors are experimentally
observed as mentioned above. The Haldane-Anderson impurity model
had been introduced to study transition-metal impurities in
semiconductors \cite{Haldane}. After the discovery of DMS, the
magnetic properties of this model were addressed within the
Hartree-Fock (HF) approximation \cite{2DFM1}, and it was shown
that long-range ferromagnetic (FM) correlations develop when the
Fermi level is located between the top of the valence band and the
impurity bound state (IBS). The FM interaction between the
impurities is mediated by the impurity-induced polarization of the
valence electron spins, which exhibit an antiferromagnetic
coupling to the impurity moments. Subsequent Quantum Monte Carlo
(QMC) calculations \cite{2DFM2} on the two-impurity
Haldane-Anderson model with the Hirsch-Fye algorithm have
supported this picture for the generation of FM correlations
between magnetic impurities in semiconductors. Various other
theoretical approaches have also been used to study magnetic
impurities in semiconductors. The Zener model has been invoked to
describe the ferromagnetism in semiconductors \cite {ZnMnO1}.
Numerical calculations based on local spin density approximation
(LSDA) have also found that magnetic states and corresponding
Curie temperatures in ZnO-based DMS are controlled by changing the
carrier density or the magnetic impurity concentration
\cite{ZnMnO2,ZnMnO3,ZnMnO4,ZnMnO5,ZnMnO6}. Within the context of
DMS, the Anderson Hamiltonian for a semiconductor host was also
considered by Krstaji\'c {\it et al.} \cite{Krstajic}, and it was
shown that an FM interaction is generated between the impurities
due to kinematic exchange. The role of IBS in producing the FM
interaction in DMS was also discussed within the ``double
resonance mechanism" using HF \cite{Inoue}.

In this paper, we present QMC results for the compound (Zn,Mn)O.
For ZnO host, the wurtzite structure is the most common phase, and
thus almost all experiments for ZnO are focused on this structure.
But ZnO with the zincblende and the rocksalt structures are also
experimentally possible in thin films and at high pressure,
respectively \cite{StrucZnO}. The band structures of ZnO with the
wurtzite, zincblende and rocksalt structures had been already
calculated within a single set of tight-binding parameters
\cite{StrucZnO}, and we will follow these results in this paper to
study the ferromagnetism for the compound (Zn,Mn)O with
experimentally determined values for the $p$-$d$ mixing and the
onsite Coulomb repulsion \cite{pdsig}. For the doped Mn$^{2+}$
impurity, we neglect the Hund coupling among the five occupied
$3d$ orbitals, and, for simplicity, consider the $3d$ orbitals
independently. In the dilute impurity limit, the Haldane-Anderson
impurity model is invoked to describe the magnetic states of
Mn$^{2+}$ ions. The results of the numerical calculations show
that the crystal structure of the ZnO host strongly influences the
energy of the IBS and the strength of the magnetic correlations
between the impurities. In particular, for the wurtzite and
zincblende structures, we find that FM correlations with an
extended range develop at low temperatures. However, for the
rocksalt structure, FM correlations have not been observed,
because, in this case, the IBS is located at much higher
frequencies. In addition, only p-type (Zn,Mn)O is found to exhibit
FM correlations.

In this paper, our purpose is to investigate the influence of the
crystal structure of the semiconductor host on the FM correlations
between the impurities. For this purpose, we combine the
tight-binding calculations for the host band structure and the
impurity-host hybridization with the QMC simulations. Our impurity
model is simple, because we neglect the Hund couplings and
consider only one of the $3d$ orbitals at the impurity site.
However, this model is sufficient to demonstrate that the host
crystal structure can be used to control ferromagnetism in DMS.
Instead of the tight-binding approximation, the Local Density
Approximations (LDA) can be used to calculate the host band
structure and the impurity-host hybridization. In addition, it is
possible to perform QMC simulations for all five of the impurity
$3d$ orbitals with the Hund couplings. We think that this way of
combining the LDA and QMC techniques can yield accurate
predictions about ferromagnetism in DMS materials in the future.

\section{Impurity model}

In order to describe the transition-metal impurities in a ZnO
host, we use the Haldane-Anderson impurity model \cite{Haldane}
which is defined by
\begin{eqnarray}
  H &=&
  \sum_{\textbf{k},\alpha,\sigma}[\epsilon_{\alpha}(\textbf{k})-\mu]
  c^{\dag}_{\textbf{k}\alpha\sigma}c_{\textbf{k}\alpha\sigma} \notag\\
  &+&\sum_{\textbf{k},\alpha,\textbf{i},\xi,\sigma}(V_{\textbf{i}\xi\textbf{k}\alpha }
 d^{\dag}_{\textbf{i}\xi\sigma} c_{\textbf{k}\alpha\sigma}
   + H.c.) \notag\\ &+ &(E_d-\mu)\sum_{\textbf{i},\xi,\sigma}
   d^{\dag}_{\textbf{i}\xi\sigma}d_{\textbf{i}\xi\sigma}
   + U\sum_{\textbf{i}}n^{\dag}_{\textbf{i}\xi\uparrow}n_{\textbf{i}\xi\downarrow},
   \label{Ham}
\end{eqnarray}
where $c^{\dag}_{\textbf{k}\alpha\sigma}$
($c_{\textbf{k}\alpha\sigma}$) is the creation (annihilation)
operator for a host electron with wavevector $\textbf{k}$ and spin
$\sigma$ in the valence ($\alpha = v$) or conduction ($\alpha = c$)
band, and $d^{\dag}_{\textbf{i}\xi\sigma}$
($d_{\textbf{i}\xi\sigma}$) is the creation (annihilation) operator
for a localized electron at impurity site $\textbf{i}$ in orbital
$\xi$ and spin $\sigma$ with
$n_{\textbf{i}\xi\sigma}=d^{\dag}_{\textbf{i}\xi\sigma}d_{\textbf{i}\xi\sigma}$.
Here, $\epsilon_{\alpha}(\textbf{k})$ is the host band dispersion,
$\mu$ the chemical potential, $V_{\textbf{i}\xi\textbf{k}\alpha}$
the mixing between the impurity and host, $E_d$ the impurity
$\xi$-level energy, and $U$ the onsite Coulomb repulsion for the
impurity.

The energy bands $\epsilon_{\alpha}(\textbf{k})$, and the
impurity-host hybridization $V_{\textbf{i}\xi\textbf{k}\alpha}$ will
be calculated within the tight-binding approximation for the
wurtzite, zincblende and rocksalt crystal structures of the ZnO host
material. For the compound (Zn,Mn)O, the value of the onsite Coulomb
repulsion for Mn$^{2+}$ is taken as $U=5.2$eV by comparing with the
photoemission spectroscopy measurements \cite{pdsig}. In addition,
because the experimental value of $E_d$ for Mn$^{2+}$ in ZnO host is
unknown, in the following we use $E_d=\mu-U/2$ so that the impurity
sites develop large magnetic moments. The results on the magnetic
correlations between the impurities depend weakly on small
variations on the value of $E_d$.

We note that, in Eq.~(1), the Hund couplings among the different
impurity $3d$ orbitals is neglected. In this paper, we consider
only one of the $3d$ orbitals at the impurity site, since here our
purpose is to demonstrate the effects of the host crystal
structure on the FM correlations. Multi-orbital effects, where we
keep all of the impurity $3d$ orbitals and the Hund couplings,
will be studied with QMC in a separate paper.

\section{Tight-binding approach for the $ZnO$ band structure
and the impurity-host hybridization}

In this section, we discuss the tight-binding calculation of the
band-structure of the ZnO host, and the impurity-host
hybridization. The energy bands, $\epsilon_{\alpha}(\textbf{k})$,
of ZnO had been already calculated for the wurtzite, zincblende
and rocksalt structures using a single set of $sp^3$ tight-binding
parameters \cite{StrucZnO}. In this approach, the basis consists
of one $4s$ and three $4p$ orbitals for cation Zn$^{2+}$ and three
$2p$ orbitals for anion O$^{2-}$. The values of the orbital
energies are $E_p$(O) = 0.550 eV, $E_s$(Zn)=3.450 eV, $E_p$(Zn) =
13.050 eV. In addition, the mixing values between the $s$, $p$
orbitals of Zn$^{2+}$ and the $p$ orbitals of O$^{2-}$ are taken
to be $(sp\sigma)$ = 2.965 eV, $(pp\sigma)$ = 4.324 eV, and
$(pp\pi)$ = -1.157 eV.

\begin{figure}[tbp]
\includegraphics[width = 8.5 cm]{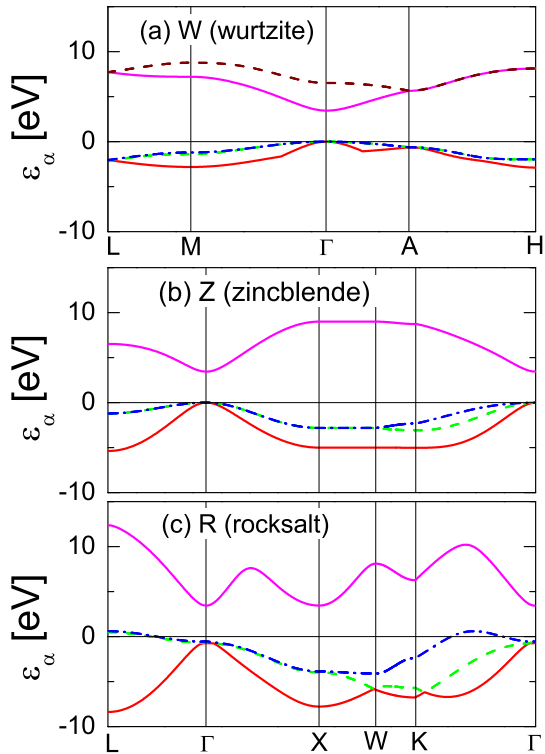}
\caption{Energy bands of the ZnO host near the semiconductor gap
with (a) wurtzite, (b) zincblende, and (c) rocksalt crystal
structures obtained using the tight-binding approximation. These
results were reproduced using the parameters given in
Ref.~\cite{StrucZnO}.} \label{F-Eig}
\end{figure}

Using these tight-binding parameters and keeping all of the branches
within the $sp^3$ basis, we have reproduced the band structure of
ZnO. In Figs. \ref{F-Eig}(a)-(c), we have plotted the branches near
the semiconductor gap. Here, we observe that, for the wurtzite and
zincblende structures, the top of the valence band is located at the
$\Gamma$ point with a direct gap of 3.45 eV. For the rocksalt case,
the top of the valence band, located at the $L$ point, is at 0.6 eV,
while the $\Gamma$ point is at -0.53 eV. Hence, for the rocksalt
structure the system has an indirect semiconductor gap.

Next, we discuss the calculation of the impurity-host
hybridization within the tight-binding approximation. Once a
substitutional impurity Mn$^{2+}$ is introduced and takes the
position of a Zn$^{2+}$ cation, the $3d$ orbital $\xi$ of
Mn$^{2+}$ will mix with the neighboring $2p$ orbitals of O$^{2-}$.
The mixing matrix element
$V_{\textbf{i}\xi\textbf{k}\alpha}\equiv\langle\varphi_{\xi}
(\textbf{i})|H|\Psi_{\alpha}(\textbf{k})\rangle$ has the form of
\begin{eqnarray}
V_{\textbf{i}\xi\textbf{k}\alpha } &=& \frac{1}{\sqrt{N}}e^{i
\textbf{k}\cdot \textbf{i}}\sum_{p,\textbf{n}}e^{i \textbf{k}\cdot
(\textbf{n}-\textbf{i})}a_{\alpha p}(\textbf{k})
\langle\varphi_{\xi}(\textbf{i})|H|\varphi_{p}(\textbf{n})\rangle
\notag\\
& \equiv &\frac{1}{\sqrt{N}}e^{i \textbf{k}\cdot
\textbf{i}}V_{\xi\alpha }(\textbf{k}),\label{E-Mix}
\end{eqnarray}
where $\varphi_{\xi} (\textbf{i})$ is the impurity $3d$-state at
site $\textbf{i}$, and $\Psi_{\alpha}(\textbf{k})$ is the host
state with wavevector $\textbf{k}$ and band index $\alpha$, which
is expanded by atomic orbitals $\varphi_{p}(\textbf{n})$ with
orbital index $p$ and site index $\textbf{n}$. Here, $N$ is the
total number of host lattice sites, and $a_{\alpha p}(\textbf{k})$
is an expansion coefficient. For the $p$-$d$ mixing integrals of
$\langle\varphi_{\xi}(\textbf{i})|H|\varphi_{p}(\textbf{n})\rangle
$, $p$ represents the three $2p$ orbitals of O$^{2-}$, and $\xi$
denotes the five $3d$ orbitals of Mn$^{2+}$. As shown by Slater
and Koster \cite{TBSK}, these fifteen mixing integrals can be
expressed by only two integrals $(pd\sigma)$ and $(pd\pi)$ and
direction cosines $l$, $m$ and $n$ in the two-center
approximation. In this approach, the integrals $(pd\sigma)$ and
$(pd\pi)$ are taken as fitting parameters which can then be
determined by other exact results or by comparison with the
experimental data. For the compound (Zn,Mn)O, the value of
$(pd\sigma)$ = -1.6 eV has been estimated by comparing with the
photoemission spectroscopy measurements \cite{pdsig}, while the
value of $(pd\pi)$ is always determined by the relation
$(pd\pi)=-(pd\sigma)/2.16$ \cite{pdpi}. We will use these values
in the rest of this paper.

It is established that, for the wurtzite or zincblende crystal
structures, the $p$-$d$ mixing is dominated by the occupied
$t_{2g}$ ($xy$, $yz$, $zx$) orbitals because of the tetrahedral
crystal field, while for the rocksalt structure, the $p$-$d$
mixing comes mainly from the occupied $e_g$ ($x^2-y^2$, $z^2$)
orbitals due to the octahedral crystal field \cite{5dorb}. In
addition, the Hund coupling among the $3d$ orbitals of Mn$^{2+}$
is neglected in this paper. In addition, for simplicity, here we
consider only one of the $t_{2g}$ orbitals ($\xi$=xy orbital here)
for the (Zn,Mn)O with wurtzite and zincblende structures, and only
one of the $e_{g}$ orbitals ($\xi=x^2-y^2$ orbital here) for the
(Zn,Mn)O with rocksalt structure. Hence, we treat the $3d$
orbitals of Mn$^{2+}$ independently. In this paper, we are mainly
interested in the effects of the host crystal structure, and we
will treat the multi-orbital effects in a separate paper.

\begin{figure}[tbp]
\includegraphics[width = 8.5 cm]{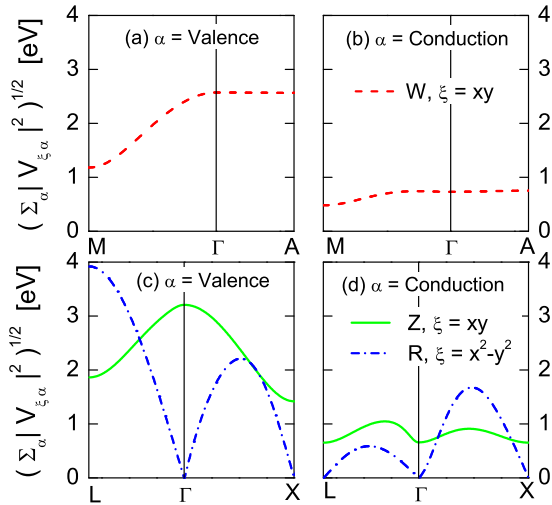}
\caption{ Hybridization function $\overline{V}_{\xi}(\textbf{k})$
of a single Mn$^{2+}$ orbital with the valence bands ((a) and (c))
or the conduction bands ((b) and (d)). Here, results are shown for
a Mn$^{2+}$ $3d(xy)$ orbital in the wurtzite or zincblende ZnO and
for a Mn$^{2+}$ $3d(x^2-y^2)$ orbital in the rocksalt ZnO. In
obtaining these results, we have used mixing parameters with
values taken from Ref. \cite{pdsig}.} \label{F-Mix}
\end{figure}

Figure \ref{F-Mix} displays results on the the $p$-$d$ mixing
function $\overline{V}_{\xi}(\textbf{k})$ defined by
\begin{equation}
\overline{V}_{\xi}(\textbf{k})  \equiv \big(
\sum_{\alpha}|V_{\xi\alpha}(\textbf{k})|^2 \big)^{1/2}
\label{Vbar}
\end{equation}
where only one of the Mn$^{2+}$ $3d$ orbitals, labeled by $\xi$,
is considered. In Eq.~(\ref{Vbar}), the summation over $\alpha$ is
performed only over the valence bands (Fig. 2(a) and Fig. 2(c)) or
the conduction bands (Fig. 2(b) and Fig. 2(d)). Here,
$\overline{V}_{\xi}(\textbf{k})$ is plotted along various cuts in
the Brillouin zone for wurtzite, zincblende and rocksalt crystal
structures. Figures 2(a) and (b) show $\overline{V}$ for a Mn
$3d(xy)$ orbital when the ZnO has the wurtzite structure. Here, we
observe that, at the $\Gamma$ point, the total hybridization of
the $xy$ orbital with the valence bands is about three times
larger than that with the conduction bands. For the wurtzite and
the zincblende structures, the semiconductor gap edges are located
at the $\Gamma$ point, hence the value of $\overline{V}$ near
$\Gamma$ will be particularly important in determining the energy
of the IBS and the strength of the magnetic correlations between
the impurities. Figures 2(c) and (d) show that, for the case of a
Mn $3d(xy)$ orbital in ZnO with the zincblende structure, the
total hybridization with the valence bands is also stronger than
that with the conduction bands near the $\Gamma$ point. Figures
2(c) and (d) also show results for a Mn $3d(x^2-y^2)$ orbital in
the rocksalt structure. Here, we see that the total hybridization
with the valence and the conduction bands vanish at the $\Gamma$
point. However, at the $L$ point, where the top of the valence
band is located, the total hybridization with the valence band is
the stronger than in the rest of the cases.

The results on the frequency of the IBS and the strength of the FM
correlations depend sensitively on the value of hybridization near
the gap edges, which we will discuss in the next section. We note
that, in turn, the hybridization depends strongly on the values of
the mixing parameters $(pd\sigma)$ and $(pd\pi)$ of the
tight-binding approach.

\section{Quantum Monte Carlo Results}

In this section, we present results on the impurity magnetic
correlations, which were obtained using the Hirsch-Fye QMC
technique \cite{QMC}. The input parameters for the QMC simulations
were calculated with the tight-binding approach described above.
The following results were obtained with more than 10$^{5}$ Monte
Carlo sweeps and Matsubara time step $\Delta\tau=0.225$.

We first discuss the local moment formation at an impurity
orbital. For this purpose, we have performed QMC simulations to
calculate $\langle (M^z)^2 \rangle$, where
\begin{equation}
M^z = n_{\textbf{i}\xi\uparrow} - n_{\textbf{i}\xi\downarrow}
\end{equation}
is the magnetization operator for a single $3d(\xi)$ orbital at
the impurity site $\textbf{i}$. We have performed the calculations
for a single $3d$ orbital because in this paper we are mainly
interested in the effects of the host crystal and we neglect the
multi-orbital effects, which we will treat in a separate paper.
Hence, $\langle (M^z)^2 \rangle$ represents the square of the
local-moment for a single $3d$ orbital added to the ZnO host.

Figure 3(a) and (b) shows results on $\langle (M^z)^2 \rangle$ for
a $3d(xy)$ orbital in wurtzite and zincblende structures and a
$3d(x^2-y^2)$ orbital in the rocksalt ZnO. In Fig. 3(a), $\langle
(M^z)^2 \rangle$ versus the chemical potential $\mu$ is plotted
for $0 < \mu < 0.35$eV and in Fig. 3(b) for $0 < \mu < 4$eV. These
results are for temperature $T=100K$. In these figures, we observe
discontinuities at $0.12$eV, $0.20$eV and $1.6$eV  for the
zincblende, wurtzite and rocksalt structures, respectively.

According to the Hartee-Fock and QMC calculations, the presence of a
discontinuity in $\langle (M^z)^2\rangle$ versus $\mu$ implies the
existence of an IBS at this energy \cite{2DFM1,2DFM2}. A step
discontinuity develops in the magnitude of the magnetic moment as
$\mu$ increases through $\omega_{IBS}$, because impurity spectral
weight is induced in the semiconductor gap at $\omega_{IBS}$ for
sufficiently strong hybridization. When the IBS is occupied, the
spin polarization of the host split-off state, which is due to the
impurity-host mixing and at the same energy as the IBS, will cancel
the spin polarization of the valence band. Thus antiferromagnetic
couplings between the polarized host carriers and the impurities
disappear. This causes the FM interaction between the impurities,
which is mediated by the polarized carriers around the magnetic
impurities, to vanish. In this paper, we are studying how this
mechanism of ferromagnetic correlations is influenced by the crystal
structure of the the ZnO host. The variation in the values of
$\omega_{IBS}$ for different crystal structures seen in Fig. 3 is
clearly a consequence of the differences in the energy bands and the
impurity-host mixing. In addition, the values of the hybridization
with the bottom of the conduction bands are weaker, and hence we do
not observe bound states near the bottom of the conduction band.

The value of $\omega_{IBS}$ plays an important role in determining
the strength of the FM correlations which develop between the
impurities when the IBS is unoccupied. Within the Hartree-Fock
approximation and for a semiconductor host with constant density
of states $\rho_0$ and semi-infinite bands \cite{2DFM1}, the range
of the ferromagnetic (FM) correlations between the impurities is
given by
\begin{equation}
\ell_0 \approx \frac{1}{\sqrt{16\pi\rho_0\omega_{IBS}}}
\label{l0}
\end{equation}
when the IBS is unoccupied ($0 < \mu < \omega_{IBS}$). However,
when the IBS becomes occupied ($\omega_{IBS} < \mu$), the FM
correlations become weaker. The QMC calculations performed for a
two-dimensional (2D) semiconductor host with quadratic
quasiparticle dispersion confirm this picture \cite{2DFM2}. These
QMC calculations show that, in 2D, $\omega_{IBS}$ increases as the
strength of the hybridization grows. The maximum range of the FM
correlations decreases as $\omega_{IBS}$ increases in agreement
with the HF calculations. However, in the three-dimensional case,
the IBS does not exist, if the hybridization is smaller than a
critical value \cite{Tomoda}.

In the following calculations, we will see that the magnetic
correlations between the impurities are sensitive to the location
of the chemical potential with respect to the IBS. An IBS with
shallower position ($\omega^{Z}_{IBS}\approx 0.1$ eV) is obtained
for (Zn,Mn)O with the zinblende structure, while a much deeper IBS
($\omega^{R}_{IBS}\approx 1.6$ eV) is found for the rocksalt case.
Based on the previous HF and QMC calculations \cite{2DFM1, 2DFM2},
it is reasonable to expect that the FM correlation range for
(Zn,Mn)O with the wurtzite structure will be shorter than that
with the zincblende structure, and will be much longer than that
of the rocksalt case.

In addition, it should be pointed out that the IBS for (Zn,Mn)O
with wurtzite and zincblende structures only exists near the top
of valence band, while no IBS is found near the bottom of
conduction band. If, instead of Mn$^{2+}$, another transition
metal ion TM$^{2+}$ is substituted into the ZnO host, then the
only different quantities would be the $p$-$d$ mixing parameters
($pd\sigma$) and ($pd\pi$) within this framework. In particular,
for various hosts and transition-metal impurities, the $p$-$d$
mixing parameters, which are consistent with the experimental
measurements, are mostly in the range $-1.6$ eV$\leq (pd\sigma)
\leq$-1.0 eV \cite{pdsig,pdsig2}, while $(pd\pi)=-(pd\sigma)/2.16$
\cite{pdpi}. We have checked the IBS for (Zn,TM)O for various
$p$-$d$ mixing values in the above mentioned range. We find that,
when smaller values are used for the $p$-$d$ mixing, the IBS
shifts towards the top of valence band, while at the same time no
IBS develops near the bottom of the conduction band. Since the
doping of the transition metal ion TM$^{2+}$ into the ZnO host
does not itself introduce carriers, the hole and electron carriers
in (Zn,TM)O are associated with \emph{additional} the acceptor or
donor defects, respectively. This implies that for p-type
(Zn,TM)O, the FM correlations will develop when $0\leq \mu \leq
\omega_{IBS}$, while for the n-type (Zn,TM)O, no FM correlations
are expected due to the absence of the IBS. In fact, the existence
of p-type rather than n-type (Zn,Mn)O with wurtzite structure has
recently been pointed out based on the analysis of the
experimental measurements \cite{PRoomT4,PRoomT5}.

\begin{figure}[tbp]
\includegraphics[width = 8.5 cm]{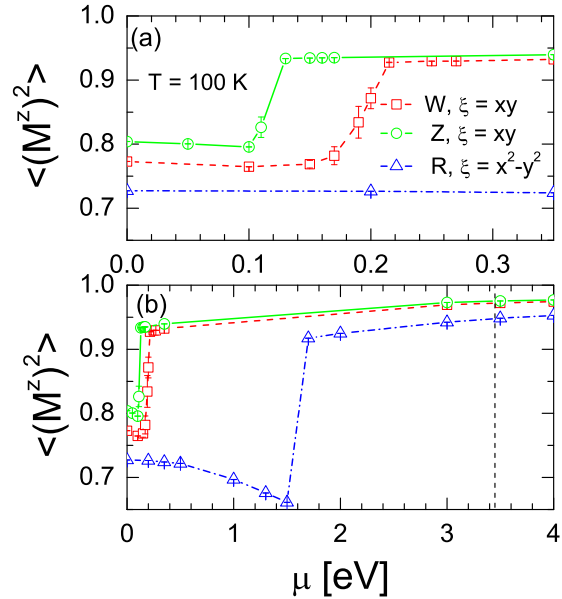}
\caption{ Square of the magnetic moment at the impurity site,
$\langle(M^z)^2\rangle$, versus the chemical potential $\mu$ for
different energy intervals at $T=100K$. These results were
obtained for a single Mn$^{2+}$ $3d(\xi)$ orbital added to the ZnO
host. For the wurtzite and zincblende crystal structures, we have
considered a $\xi=xy$ orbital, while for the rocksalt case an
$x^2-y^2$ orbital. The vertical dashed line denotes the bottom of
the conduction band. } \label{F-C1}
\end{figure}

\begin{figure}[tbp]
\includegraphics[width = 8.5 cm]{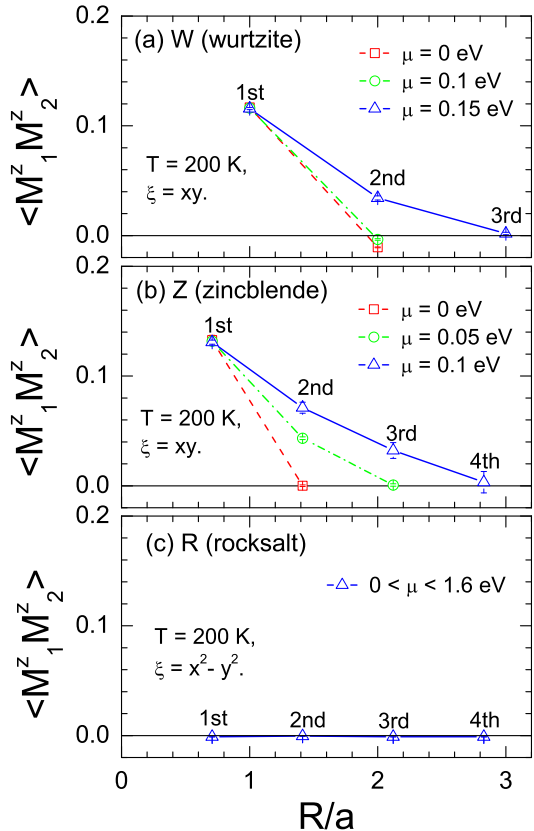}
\caption{ Impurity-impurity magnetic correlation function $\langle
M^z_{1}M^z_{2}\rangle$ versus distance $R/a$ for (a) the wurtzite,
(b) zincblende, and (c) rocksalt structures. Here, each impurity
site consists of a single Mn$^{2+}$ $3d$ orbital. For the wurtzite
and zincblende crystal structure, we have considered a $3d(xy)$
orbital at the impurity sites, while for the rocksalt case a
$3d(x^2-y^2)$ orbital. } \label{F-C2}
\end{figure}

We next display results on the impurity-impurity magnetic
correlation function $\langle M^z_{1}M^z_{2}\rangle$ versus the
impurity separation $R/a$ in Fig. \ref{F-C2}. These results are
for two Mn$^{2+}$ $3d$ orbitals, which are of the same type, added
to the ZnO host. Clearly, in a more realistic calculation of FM
correlations between the impurities, it would be necessary to
consider the magnetic correlations between different types of $3d$
orbitals. However, our purpose in this paper is to investigate the
role of the host electronic structure. In Fig. 4, the temperature
$T = 200K$, $R=|R_1-R_2|$, and $a$ is the lattice constant. In our
chosen directions, the distance between the
1st-nearest-neighboring Mn$^{2+}$ is $a$ for the wurtzite
structure, while it is $a\sqrt{2}/2$ for the zincblende and
rocksalt structures. As shown in Fig. \ref{F-C2}(a) for the
wurtzite structure, the impurity spins exhibit FM correlations at
$\mu=0.0$, but it is short range. By increasing $\mu$ to 0.15 eV,
the range of the FM correlations becomes longer extending to the
3rd-nearest neighboring Mn$^{2+}$. Further increasing $\mu$, the
FM correlations becomes weaker. This is because the IBS of
(Zn,Mn)O with wurtzite structure becomes occupied as $\mu$ changes
above 0.15 eV, as seen in Fig. \ref{F-C1}. For the zincblende
structure, similar results are obtained as shown in Fig.
\ref{F-C2}(b): the impurity spins exhibit FM correlations
extending to the 1st-nearest neighboring Mn$^{2+}$ at $\mu=0.0$.
Increasing $\mu$ to 0.1 eV, the FM correlations extend to the
4th-nearest neighboring Mn$^{2+}$. Further increasing $\mu$ to
above 0.1 eV, the FM correlations become weaker. This is because
the IBS of (Zn,Mn)O with the zincblende structure becomes occupied
as $\mu > 0.1$ eV, as displayed in Fig. \ref{F-C1}. We note that
the compound (Zn,Mn)O with the zincblende structure seems to
possess a longer range for the FM correlations than that with the
wurtzite structure, while almost all existing experiments for ZnO
are focused on the wurtzite structure. Not surprisingly, we have
not observed magnetic correlations for the rocksalt structure in
Fig. \ref{F-C2}(c). This is because the position of IBS for
(Zn,Mn)O with rocksalt structure is too deep as seen in Fig.
\ref{F-C1}, thus the FM correlation range \cite{2DFM1, 2DFM2} is
shorter than the first nearest-neighbor distance.

We next study the temperature dependence of the ferromagnetic
correlations between the impurities in different crystal
structures. Figure 5 displays the $T$ dependence of $\langle
M^z_{1}M^z_{2}\rangle$ for impurities which are second-nearest
neighbors in wurtzite, zincblende and rocksalt crystal structures.
For wurtzite and zincblende structures, these calculations have
been performed for the $3d(xy)$ orbitals, and for the rocksalt
case they are for the $3d(x^2-y^2)$ orbital. In addition, for the
wurtzite and zincblende structures, the chemical potential is
taken as $\mu=0.15$ eV and 0.1 eV, respectively, where the IBS is
unoccupied and the longest range of the FM correlations occurs as
shown in Fig. \ref{F-C2}. The results for the rocksalt case are
for $\mu=1.5$eV, so that the IBS is unoccupied here also. We
observe that, as $T$ decreases from $800K$ down to $200K$, the FM
correlations grow for the wurtzite and zincblende structures. In
this $T$ range, the FM do not develop for the rocksalt case.

\begin{figure}[tbp]
\includegraphics[width = 8.5 cm]{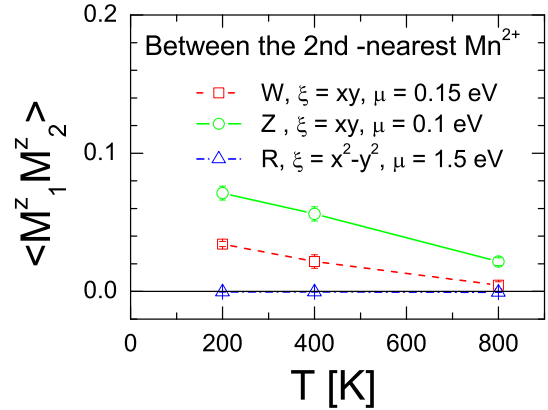}
\caption{ Impurity-impurity magnetic correlation function $\langle
M^z_{1}M^z_{2}\rangle$ for second nearest-neighbor impurity sites
versus temperature $T$. Here, for the wurtzite and zincblende
crystal structures, we have considered $3d(xy)$ orbitals, while
for the rocksalt case $3d(x^2-y^2)$ orbitals. } \label{F-Temp}
\end{figure}

\section{Discussion}

For the ZnO host with the rocksalt structure, it is useful to
mention previous studies \cite{StrucZnO,Rocksalt1,Rocksalt2} which
are related to the \emph{indirect} energy gaps obtained with the
$sp^3$ tight-binding calculation shown in Fig. \ref{F-Eig}(c).
Using the Hartree-Fock method, Jaffe et al. \cite{Rocksalt1}
pointed out that for the rocksalt ZnO, the point symmetry in
rocksalt structure does not allow for mixing between the Zn $3d$
orbitals and the O $2p$ orbitals at the $\Gamma$ point, but mixing
is allowed elsewhere in the Brillouin zone. Thus the valence band
maximum shifts away form the $\Gamma$ point, so that the gap
becomes \emph{indirect}. As mentioned by Skinner and LaFemina in
Ref.\cite{StrucZnO}, where we follow their $sp^3$ tight-binding
calculation for the ZnO host in present paper, this effect also
exists even when the $sp^3$ tight-binding model does not
\emph{explicitly} include the Zn $3d$ orbitals, because the Zn
$3d$ character has been \emph{implicitly} included in the hopping
integrals through the tight-binding interpolation of the bulk
\emph{ab} \emph{initio} pseudopotential bands \cite {Rocksalt2}
that Skinner and LaFemina used to derive them.

However, there also exist calculations for the rocksalt ZnO, which
are based on the quasiparticle approach \cite{Rocksalt3} and the
pseudopotential method \cite{Rocksalt4}, yielding a \emph{direct}
energy gap at the $\Gamma$ point. In order to understand the
results observed for (Zn,Mn)O with the rocksalt structure, we have
considered the case of MgO, whose crystal structure is also
rocksalt but has a \emph{direct} energy gap at the $\Gamma$ point
\cite{MgO}. Using the tight-binding parameters given in
Ref.~\cite{MgO}, where the host basis consists of one $s$ orbital
for Mg$^{2+}$ and three $2p$ orbitals for O$^{2-}$, we have
calculated the $p$-$d$ mixing for the compound (Mg,Mn)O as in the
case for (Zn,Mn)O. We found that, near the $\Gamma$ point, the
$p$-$d$ mixing for (Mg,Mn)O with the rocksalt structure is also
close to zero, where $d$ stands for the five $3d$ orbitals. These
results imply that the host compound with the rocksalt structure
does not mix with the doped $3d$ orbitals at the $\Gamma$ point
regardless of whether there is a direct or indirect energy gap. In
fact, this is generally true if the host with the rocksalt
structure can be described using the $sp^3$ tight-binding
approach. At the $\Gamma$ point, because of the orthogonality of
the $s$ and $p$ orbitals, we find that, for a fixed band $\alpha$,
the coefficients $a_{\alpha p}$ in Eq.(\ref{E-Mix}) have only one
non-zero value, for example, $a$. Thus the $p$-$d$ mixing at the
$\Gamma$ point has the form
\begin{eqnarray}
V_{\alpha}(0)=a\sum_{\textbf{n}}
\langle\varphi_{\xi}(\textbf{i})|H|\varphi_{p}(\textbf{n})\rangle
\equiv a\sum_{\textbf{n}}E_{\xi,p}(\textbf{i}-\textbf{n})
\label{E-Mix-0}
\end{eqnarray}
where $\xi$ represents one of the five $3d$ orbitals, $p$ denotes
one of the $s$, $p_x$, $p_y$, and $p_z$ orbitals. In addition, here,
the summation is taken over the nearest neighbors, and
$E_{\xi,p}(\textbf{i}-\textbf{n})$ is given using the notation of
Slater and Koster \cite{TBSK}, which can be expressed in terms of
the non-zero $(pd\sigma)$ and $(pd\pi)$, and the direction cosines
$l$, $m$, and $n$. For the rocksalt structure, it can be shown that
the value of $V_{\alpha})(0)$, Eq.~(\ref{E-Mix-0}), vanishes for any
$\xi$ and $p$ orbitals, when the summation is performed over the six
nearest neighbors.

It is obvious that the magnetic correlations between the impurities
depend on the position of the IBS, while the IBS is closely related
the host band structure as well as the host-impurity mixing
parameters. As mentioned above, for the rocksalt ZnO doped with
magnetic impurities, the IBS will shift to the top of valence band
if we decrease the values of the $p$-$d$ mixing parameters
$(pd\sigma)$ and $(pd\pi)$. Thus the detailed band structure
calculations, such as LDA, are needed in the future to combine with
QMC so that the ZnO-based DMS can be described more accurately.

\section{Summary and Conclusions}

In summary, we have studied the ferromagnetism for the compound
(Zn,Mn)O with different crystal structures in the dilute impurity
limit based on the Haldane-Anderson impurity model. The band
structures of the ZnO host were calculated using the $sp^3$
tight-binding parameters from Ref. \cite{StrucZnO}, and the
$p$-$d$ mixing parameters and the onsite Coulomb repulsion $U$
were obtained from comparisons with the photoemission measurements
on (Zn,Mn)O \cite{pdsig}. The QMC calculations show that the
magnetic correlations between Mn$^{2+}$ impurities in ZnO are
strongly affected by the host crystal structure. For the wurtzite
and zincblende structures, the ferromagnetic correlations are
found, and their range extends up to the third or the fourth
nearest neighbor sites at low temperatures. On the other hand, for
the rocksalt structure, no magnetism has been found even between
the nearest-neighbor impurities. In addition, only p-type ZnO
doped with magnetic impurities is found to have ferromagnetism.

In this paper, our main purpose has been to investigate the
effects of the crystal structure on the IBS and the FM
correlations for the ZnO host. For simplicity, we have neglected
the Hund coupling among the Mn$^{2+}$ $3d$ orbitals. In addition,
we have considered only one of the five $3d$ orbitals at the
impurity sites. This is clearly an over simplification, however,
it does allow us to demonstrate the role of the host crystal
structure in determining the energy of the IBS and the maximum
range of the FM correlations. Currently, we are in the process of
performing QMC calculations which include all five of the $3d$
orbitals at the impurity site without neglecting the Hund
couplings. In addition, we are using the LDA technique, instead of
the tight-binding approximation, to calculate more precisely the
host band structure and the impurity-host hybridization. We think
that this LDA+QMC approach can be used to make accurate
predictions about the FM correlations for different sets of DMS
materials.

It is generally agreed up on that the ferromagnetism in the DMS
can be controlled by changing the type of the transition-metal
impurities or the host semiconductor as well as the occupation of
the IBS. The results presented in this paper suggest that the host
crystal structure can also be used in the search for high-$T_c$
ferromagnetism in DMS.

\section*{ACKNOWLEDGMENTS} This work was supported by the NAREGI
Nanoscience Project and a Grant-in Aid for Scientific Research
from the Ministry of Education, Culture, Sports, Science and
Technology of Japan, and NEDO. The authors thank the Supercomputer
Center at the Institute for Solid State Physics, University of
Tokyo, for the use of the facilities.


\begin{thebibliography}{99}

\bibitem{GaMnAs1} H. Ohno, A. Shen, F. Matsukura, A. Oiwa, A. Endo, S. Katsumoto, and
Y. Iye, Appl. Phys. Lett. \textbf{69}, 363 (1996); H. Ohno, Science,
\textbf{281}, 951 (1998).

\bibitem{GaMnAs2} A. M. Nazmul, S. Sugahara, and M. Tanaka, Phys. Rev. B, \textbf{67},
241308(R) (2003).

\bibitem{GaMnAs3} T. Jungwirth, K. Y. Wang, J. Masek, K. W. Edmonds, J. Konig, J. Sinova, M. Polini,
N. A. Goncharuk, A. H. MacDonald, M. Sawicki, A. W. Rushforth, R. P.
Campion, L. X. Zhao, C. T. Foxon, and B. L. Gallagher, Phys. Rev. B
\textbf{72}, 165204 (2005).

\bibitem{ZnMnO1} T. Dietl, H. Ohno, F. Matsukura, J. Cibert,
and D. Ferrand, Scinece, \textbf{287}, 1019 (2000).

\bibitem{ZnMnO2} K. Sato, and H. K. Yoshida, Jpn. J. Appl. Phys. \text{40}, L334 (2001);
Physica E, \textbf{10}, 251 (2001).

\bibitem{InRoomT1} P. Sharma, A. Gupta, K. V. Rao, F. J. Owens, R. Sharma, R. Ahuja, J. M. O. Guillen,
R. Johansson, and G. A. Gehring, Nat. Mater. \textbf{2}, 673 (2003).

\bibitem{InRoomT2} M. A. Garcia, M. L. Ruiz-Gonzalez, A. Quesada, J. L. Costa-Kramer,
J. F. Fernandez, S. J. Khatib, A. Wennberg, A. C. Caballero, M. S.
Martin-Gonzalez, M. Villegas, F. Briones, J. M. Gonzalez-Calbet, and
A. Hernando, Phys. Rev. Lett. \textbf{94}, 217206 (2005).

\bibitem{InRoomT3} J. R. Neal, A. J. Behan, R. M. Ibrahim, H. J. Blythe, M. Ziese,
A. M. Fox, and G. A. Gehring, Phys. Rev. Lett. \textbf{96}, 197208
(2006).

\bibitem{PRoomT1} N. S. Norberg, K. R. Kittilstved, J. E. Amonette, R. K. Kukkadapu, D. A. Schwartz,
and D. R. Gamelin, J. Am. Chem. Soc. \textbf{126}, 9387 (2004).

\bibitem{PRoomT2} S. W. Lim, M. C. Jeong, M. H. Ham, and J. M. Myoung,
Jpn. J. Appl. Phys. \textbf{43}, L280 (2004).

\bibitem{PRoomT3} M. Ivill, S. J. Pearton, D. P. Norton, J. Kelly, and A. F. Hebard,
J. Appl. Phys. \textbf{97}, 053904 (2005).

\bibitem{PRoomT4} K. R. Kittilstved, N. S. Norberg, and D. R. Gamelin, Phys. Rev. Lett.
\textbf{94}, 147209 (2005).

\bibitem{PRoomT5} K. R. Kittilstved, W. K. Liu, and D. R. Gamelin, Nat. Mater.
\textbf{5}, 291 (2006).

\bibitem{LowT} S. W. Jung, S. J. An, G. C. Yi, C. U. Jung, S. I. Lee, and S. Cho,
Appl. Phys. Lett. \textbf{80}, 4561 (2002).

\bibitem{NoFM} G. Lawes, A. S. Risbud, A. P. Ramirez, and R. Seshadri,
Phys. Rev. B, \textbf{71}, 045201 (2005).

\bibitem{Glass} T. Fukumura, Z. Jin, M. Kawasaki, T. Shono, T.
Hasegawa, S. Koshihara, and H. Koinuma, Appl. Phys. Lett.
\textbf{78}, 958(2001).

\bibitem{Paramag} A. Tiwari, C. Jin, A. Kvit, D. Kumar, J. F. Muth, J. Narayan,
Solid State Commun. \textbf{121}, 371(2002).

\bibitem{Haldane} F.D.M. Haldane and P.W. Anderson, \prb {\bf 13}, 2553
(1976).

\bibitem{2DFM1} M. Ichimura, K. Tanikawa, S. Takahashi, G.
Baskaran, and S. Maekawa, {\it Foundations of Quantum Mechanics in
the Light of New Technology}, eds. S. Ishioka and K. Fujikawa,
(World Scientific, Singapore, 2006), 183-186, (cond-mat/0701736).

\bibitem{2DFM2} N. Bulut, K. Tanikawa, S. Takahashi, and S. Maekawa,
Phys. Rev. B \textbf{76}, 045220 (2007).

\bibitem{ZnMnO3} K. Sato, and H. K. Yoshida, Phys. Stat. Sol. (b)
\textbf{229}, 673 (2002).

\bibitem{ZnMnO4} H. K. Yoshida, and K. Sato, Physica B
\textbf{327}, 337 (2003).

\bibitem{ZnMnO5} E. C. Lee, and K. J. Chang, Phys. Rev. B \textbf{69}, 085205 (2004).

\bibitem{ZnMnO6} N. A. Spaldin, Phys. Rev. B \textbf{69}, 125201 (2004).

\bibitem{Krstajic} P.M. Krstaji\'c, V.A. Ivanov, F.M.
Peeters, V. Fleurov, and K. Kikoin, Europhys. Lett. {\bf 61}, 235
(2003).

\bibitem{Inoue} J. Inoue, S. Nonoyama, and H. Itoh, Phys. Rev.
Lett. {\bf 85}, 4610 (2000).

\bibitem{StrucZnO} A. J. Skinner and J. P. LaFemina, Phys. Rev. B \textbf{45}, 3557(1992).

\bibitem{pdsig} T. Mizokawa, T. Nambu, A. Fujimori, T. Fukumura, and M. Kawasaki,
Phys. Rev. B \textbf{65}, 085209 (2002).

\bibitem{TBSK} J. C. Slater and G. F. Koster, Phys. Rev. \textbf{94}, 1498(1954).

\bibitem{pdpi} \emph{Electronic Structure and the Properties of
Solids}, W. Harrison (Freeman, San Francisco, 1980)

\bibitem{5dorb} \emph{Physics of Transition Metal Oxides}, S. Maekawa, T. Tohyama,
S. E. Barnes, S. Ishihara, W. Koshibae, and G. Khaliulin (Springer
Series in Solid-State Sciences, Vol. 144, 2004)

\bibitem{QMC} J. E. Hirsch and R. M. Fye, Phys. Rev. Lett. \textbf{56}, 2521 (1986).

\bibitem{Tomoda} Y. Tomoda, N, Bulut, and S. Maekawa, unpublished.

\bibitem{pdsig2} J. W. Quilty, A. Shibata, J. Y. Son, K. Takubo, T. Mizokawa, H. Toyosaki,
T. Fukumura, and M. Kawasaki, Phys. Rev. Lett. \textbf{96}, 027202
(2006).

\bibitem{Rocksalt1} J. E. Jaffe, R. Pandey, and A. B. Kunz, Phys. Rev. B \textbf{43}, 14030 (1991).

\bibitem{Rocksalt2} J. R. Chelikowsky, Solid State Commun. \textbf{22}, 351 (1977).

\bibitem{Rocksalt3} H. Q. Ni, Y. F. Lu, and Z. M. Ren, J. Appl. Phys. \textbf{91}, 1339 (2002).

\bibitem{Rocksalt4} D. Fritsch, H. Schmidt, and M. Grundmann, Appl. Phys. Lett. \textbf{88}, 134104 (2006).

\bibitem{MgO} V. C. Lee, and H. S. Wong, J. Phys. Soc. Jpn. \textbf{45}, 895 (1978).

\end{thebibliography}
\end{document}